\documentclass{PoS}

\usepackage{amsmath,amssymb,subfigure,wrapfig}

\newcommand{\be}{\begin{equation}}
\newcommand{\ee}{\end{equation}}
\newcommand{\bea}{\begin{eqnarray}}
\newcommand{\eea}{\end{eqnarray}}
\newcommand{\hk}{\hspace{0.1cm}}

\newcommand{\rk}{\right)}
\newcommand{\lk}{\left(}

\def\cb{\bar{c}} 
\def\D{\mathcal{D}}

\def\fc{\frac} 
\def\Tr{\mbox{Tr}}

\newcommand{\Det}{\mbox{Det}}

\title{\begin{center}Hamiltonian Flow of Yang-Mills Theory \\in Coulomb Gauge \end{center}}

\ShortTitle{Hamiltonian Flow of Yang-Mills Theory in Coulomb Gauge}

\author{\speaker{Hugo Reinhardt}\thanks{supported by DFG-Re856/6-3}\\
Universit\"at T\"ubingen\\
        Institut f\"ur Theoretische Physik\\
Auf der Morgenstelle 14\\
D-72076 T\"ubingen\\
Germany\\
        E-mail: \email{hugo.reinhardt@uni-tuebingen.de}}

\author{Markus Leder $^a$, Jan M.~Pawlowski $^{b}$ 
        and Axel Weber $^{c}$\\
\llap{$^a$}Universit\"at T\"ubingen,
     Institut f\"ur Theoretische Physik, Auf der Morgenstelle 14, D-72076 T\"ubingen, Germany\\
\llap{$^b$}Universit\"at Heidelberg,
       Institut f\"ur Theoretische Physik, Philosophenweg 16, D-62910 Heidelberg, Germany\\
\llap{$^c$}Instituto de F\'isica y Matem\'aticas, Universidad Michoacana de San Nicol\'as de Hidalgo, Edificio C-3, Ciudad Universitaria, 58040 Morelia, Michoac\'an, Mexico\\
       }

       \abstract{A new functional renormalization group equation for
         Hamiltonian Yang-Mills theory in Coulomb gauge is presented
         and solved for the static gluon and ghost propagators under
         the assumption of ghost dominance. The results are compared
         to those obtained in the variational approach.}

\FullConference{The XXVIII International Symposium on Lattice Field Theory, Lattice2010\\
		June 14-19, 2010\\
		Villasimius, Italy}

\begin{document}

\section{Introduction}

Although lattice calculations have given us much insight into the
non-perturbative regime of QCD, a thorough understanding of infrared
phenomena like confinement and chiral symmetry breaking has not been
achieved yet by lattice calculations as even the largest lattice
calculations presently available do not yet probe sufficiently deep
into the infrared. Hence, in recent years there have been many
activities devoted to the non-perturbative studies of the infrared
sector of continuum QCD. Among these are functional renormalization
group (FRG) equations \cite{Leder:2010ji,Litim:1998nf} and the variational approach
to the Hamiltonian formulation of QCD \cite{Szczepaniak:2001rg,Feuchter:2004mk}.  Each approach has its own advantages and
drawbacks and by combining these approaches one can expect to gain new
insights into the theory, in particular into the non-perturbative
regime.

My talk is devoted to the application of FRG flows to the Hamiltonian
formulation of Yang-Mills theory as put forward in
\cite{Leder:2010ji}. First, to motivate the FRG studies, I
will briefly summarize some essential results of the variational
Hamiltonian approach. Then I will give a short general introduction to
FRG equations and present the new flow equations for Hamiltonian
Yang-Mills theory in Coulomb gauge together with their solutions for
the gluon and ghost propagators \cite{Leder:2010ji}.

\section{Summary of the variational approach to Hamiltonian Yang-Mills theory}

\begin{figure}[t]
\centering
\parbox{.45\linewidth}{
\includegraphics[width=\linewidth]{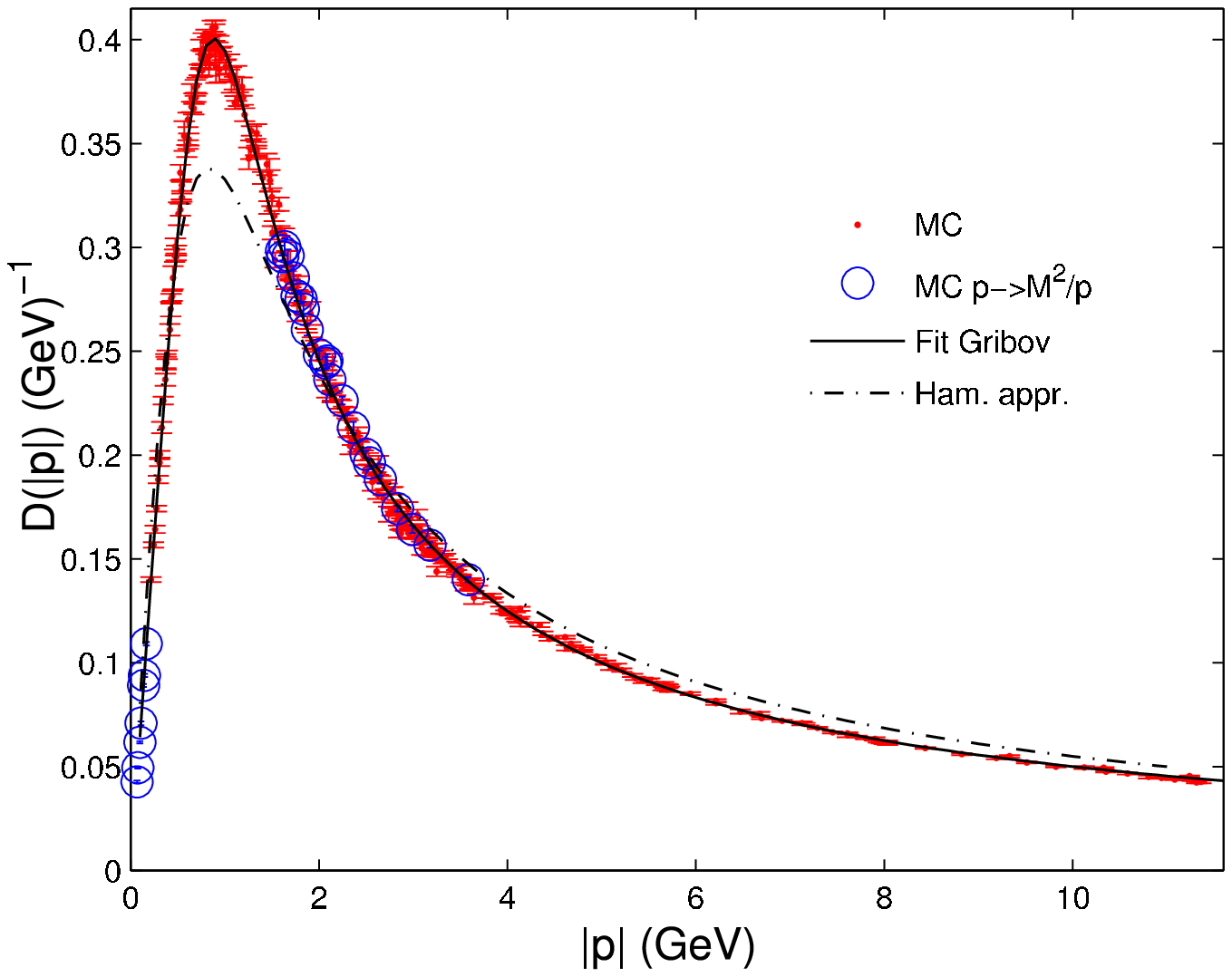}
\caption{Gluon propagator from the variational approach compared to lattice data}
\label{comp-gluon-prop-lat-var.ps}
}
\hfill
\parbox{.45\linewidth}{
\includegraphics[width=\linewidth]{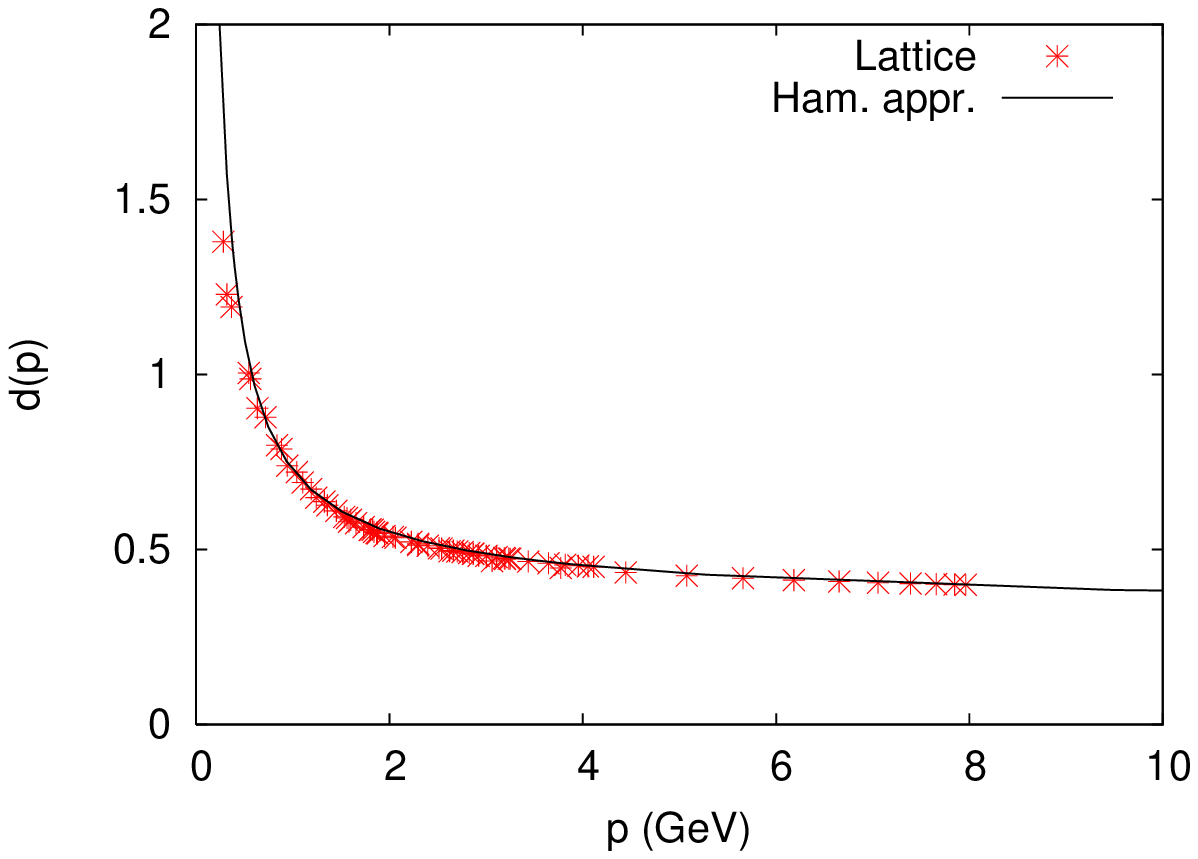}
\caption{Ghost form factor from the variational approach compared to lattice data}
\label{comp-ghost-prop-lat-var.eps}
}
\end{figure}

In the variational approach to Yang-Mills theory in Coulomb gauge
developed in our group one makes the following ansatz for the vacuum
wave functional \cite{Feuchter:2004mk}: \be
\label{1}
\psi = \frac{1}{\sqrt{\Det (- D \partial)}} \exp \lk - \frac{1}{2}
\int A \omega A \rk \hk , \ee where $\Det (- D \partial)$ is the
Faddeev-Popov determinant and $\omega$ is a variational kernel, which
is related to the static gluon propagator by \be
\label{2}
\langle A A \rangle = (2 \omega)^{- 1} \ee and therefore has the
meaning of the gluon energy. Minimizing the vacuum energy density with
respect to $\omega (p)$ \cite{Feuchter:2004mk} one finds the gluon
propagator \cite{epple}, shown in
Fig. \ref{comp-gluon-prop-lat-var.ps} together with lattice data
\cite{Burgio:2009xp}.
Both, the lattice data and the continuum solution can be fitted with
Gribov's formula \be
\label{5}
\omega (p) = \sqrt{\frac{M^4}{p^2} + p^2} \ee with $M \simeq 860 \hk
MeV$.  At large momenta $\omega (p) \sim p$, in accordance with
asymptotic freedom, whereas at small momenta $\omega (p) \sim 1/p$,
and this divergence shows the absence of gluons from the physical
spectrum, also called gluon confinement.
Fig. \ref{comp-ghost-prop-lat-var.eps} shows the ghost form factor $d
(p)$, defined by \be
\label{6}
\langle (- D \partial)^{- 1} \rangle = \frac{d (p)}{p^2} \;, \ee
obtained in the variational approach and on the lattice. In the
variational approach one can show that the infrared exponents of the
ghost and gluon propagators, \be
\label{7}
\omega (p\to 0) \sim 1/p^\alpha \hk , \qquad \hk d (p\to 0) \sim
1/p^\beta \hk , \ee are related by a sum rule under the assumption of
a trivial scaling of the ghost-gluon vertex
\cite{Schleifenbaum:2006bq}, 
\be
\label{8}
\alpha = 2 \beta - 1 \hk . \ee 
Furthermore, assuming an infrared
scaling behaviour of the ghost form factor, i.e. $\beta > 0$, one
finds two ``critical'' solutions \be
\label{9}
i)  \quad \alpha = 0.6 \,,\quad \beta = 0.8 \qquad
ii) \quad \alpha = 1 \,,\quad\ \beta = 1  \,.
\ee
\begin{wrapfigure}{R}{0.48\linewidth}
\includegraphics[width=\linewidth]{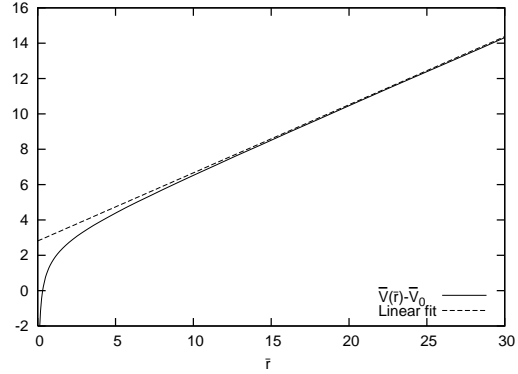}
\caption{Static quark potential}\label{quark-ir.eps}
\end{wrapfigure}
The latter results in a linearly rising quark potential shown in
Fig. \ref{quark-ir.eps}. However, sub-critical solutions $d^{- 1} (0)
\neq 0$ do also exist \cite{Epple:2007ut}. Previous lattice
calculations in Coulomb gauge seemed to yield $\beta =
\frac{1}{2}$. However, these calculations were carried out on rather
small lattices. As one increases the lattice size, $\beta$ seems to
increase as well and recent lattice calculations \cite{Burgio:2009xp}
are at least compatible with $\beta= 1$.  In Coulomb gauge the inverse
ghost form factor has been shown to represent the dielectric function
of the Yang-Mills vacuum \cite{Reinhardt:2008ek}, $\epsilon (p) = d^{-
  1} (p)$, and the so-called horizon condition $d^{- 1} (0) = 0$
implies then that the Yang-Mills vacuum is a perfect dual color
superconductor.

The advantage of the Hamiltonian formulation is its close connection
to physics. In the variational approach one makes an ansatz for the
unknown vacuum wave functional which encodes all the physics. This
ansatz can be systematically improved towards the full theory. The
price to pay is the apparent loss of renormalization group invariance
which already complicates the discussion of perturbative
renormalization, but also the discussion of scaling laws in the infrared. 

\section{FRG flow equation}
Renormalization group invariance is naturally built-in in the
functional renormalization group approach to the Hamilton formulation
of Yang-Mills theory which hence has the advantage of combining
renormalization group invariance with the physical Hamiltonian
picture.

The starting point of the FRG flow equation approach
\cite{Litim:1998nf} is the (renormalized) generating functional of
Green's functions \be
\label{11}
Z [j] = \int \D \varphi \, e^{- S [\varphi] + j \cdot \varphi} \hk , \ee
where $\varphi$ and $j$ denote collectively all fields and sources. In
the FRG approach $Z [j]$ is infrared regulated by adding the regulator
term \be
\label{12}
\Delta S_k [\varphi] = \frac{1}{2} \varphi \cdot R_k \cdot \varphi
\equiv \frac{1}{2} \int \varphi R_k \varphi
\ee
to the classical action,
\be
\label{14}
Z_k [j] = \int \D \varphi \, e^{- S [\varphi] - \Delta S_k [\varphi] + j
  \cdot \varphi} \equiv e^{W_k [j]} \hk .  \ee The regulator function
$R_k (p)$ is an effective momentum dependent mass with the
properties \be
\label{13}
\lim\limits_{p/k \to 0} R_k (p) > 0 \hk , \hk \lim\limits_{k/p \to 0} R_k
(p) = 0 \hk , \ee 
which ensures that $R_k (p)$ suppresses propagation of modes
with $p \lesssim k$ while those with $p \gtrsim k$ are unaffected and
the full theory at hand is recovered as the cut-off scale $k$ is
pushed to zero.

By taking the derivative of Eq. (\ref{14}) with respect to the
(dimensionless) momentum cut-off $t = \ln k /k_0$, one obtains the
flow equation for the generating functional of connected Green's
functions
\be
\label{15}
\partial_t W_k[j]  =  - \frac{1}{2} \frac{\delta W_k}{\delta j} \dot{R}_k 
\frac{\delta W_k}{\delta j}  - \frac{1}{2} \Tr \dot{R}_k \frac{\delta^2 W_k}{\delta j \delta j} \hk 
\ee
(where the dot denotes $\partial_t$).  In practice, it is usually more
convenient to perform first the Legendre transform from the source $j$
to the classical field 
\be
\label{16}
\phi = \frac{\delta W_k [j]}{\delta j}
\ee
resulting in the effective action
\be
\label{17}
\Gamma_k [\phi] = \lk - W_k [j] + j \cdot \phi 
\rk_{j = j_k [\phi]} - \frac{1}{2} \phi \cdot R_k \cdot \phi \hk ,
\ee
where $j_k [\phi]$ is given by solving Eq. (\ref{16}) for $j$. By
taking the derivative of Eq. (\ref{17}) w.r.t. $t = \ln k /k_0$ and
using Eq. (\ref{15}) one derives Wetterich's flow equation for
$\Gamma_k$, \be
\label{18}
\partial_t \Gamma_k [\phi] = \frac{1}{2} \Tr\, 
\frac{1}{ \Gamma_k^{(2)} [\phi] + R_k } \,\dot{R}_k  ,
\ee
where
\be
\label{19}
\Gamma^{(n)}_{k, 1 \dots n} [\phi] = \frac{\delta^n 
\Gamma_k[\phi]}{\delta \phi_1 \dots \delta \phi_n}
\ee
are the one-particle irreducible $n$-point functions (proper
vertices). The generic structure of the flow equation (\ref{18}) is
independent of the details of the underlying theory, i.e., of the
explicit form of the action $S [\phi]$, but is a mere consequence of
the form of the regulator term (\ref{12}), i.e., that it is quadratic
in the field. By taking functional derivatives of Eq. (\ref{18}) one
obtains the flow equations for the (inverse) propagators and proper
vertices. For the two-point function this equation reads \be
\label{20}
\partial_t {\Gamma}^{(2)}_{k, 12} = \frac{1}{2} \Tr\, \dot{R}_k\, \frac{1}{\Gamma_k^{(2)} + R_k} 
\left\{ - \Gamma^{(4)}_{k, 12} + \left[ \Gamma^{(3)}_{k, 1} \,\frac{1}{\Gamma_k^{(2)} + R_k} 
\,\Gamma^{(3)}_{k, 2} + \lk 1 \leftrightarrow 2 \rk \right] \right\} \frac{1}{\Gamma^{(2)}_k + R_k} ,
\ee
where all cyclic indices (summed over in the trace) have been suppressed.

\section{Hamiltonian flow}

In the Hamiltonian approach the generating functional of static correlation functions reads 
\be
\label{21}
Z [j] = \langle \psi | \exp (j \cdot \varphi) | \psi \rangle = \int \D \varphi | \psi [\varphi] |^2 \exp (j \cdot \varphi) \hk ,
\ee
where $\langle \varphi | \psi \rangle = \psi [\varphi]$ is the vacuum wave functional. With the identification 
\be
\label{22} 
| \psi [\varphi] |^2 \equiv \exp (- S [\varphi]) \;,
\ee
defining the classical action $S [\varphi]$ in terms of the vacuum
wave functional $\psi [\varphi]$, Eq. (\ref{21}) has precisely the
standard form of a generating functional except that the functional
integral extends over time independent fields.  Here we are interested
in the Hamiltonian flow of Yang-Mills theory in Coulomb gauge whose
generating functional reads
\be
\label{23}
Z [J] = \int \D A \, \Det (- D \partial) | \psi [A] |^2 \exp (J \cdot A) \hk ,
\ee
where the integration is over transversal gauge fields $A$ and the
Coulomb gauge condition has been implemented by the usual
Faddeev-Popov method. Representing the Faddev-Popov determinant in the
standard fashion by ghost fields, $c, \cb$, 
\be
\label{264}
\Det (- D \partial) = \int \D \bar{c} \D c \, e^{- \int \bar{c} (- D \partial) c}
\ee
the underlying action, cf. Eq. (\ref{22}), reads
\be
\label{24}
S [\varphi] = - \ln | \psi [A] | + \int \bar{c} (- D \partial) c \hk .
\ee 
The general flow equation (\ref{20}) still holds provided that
$\phi$ is interpreted as the superfield $\phi = (A, c, \bar{c})$. The FRG flow
equations for the gluon and ghost propagators are diagrammatically
given in Fig. \ref{full flow}.

\section{Approximation schemes and numerical solution}

\begin{figure*}[h] 
\centering
\large{$k\partial_k$} \parbox{30pt}{\includegraphics[width=\linewidth]{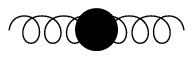}}{\large$^{-1}=$} 
\parbox{50pt}{\includegraphics[width=\linewidth]{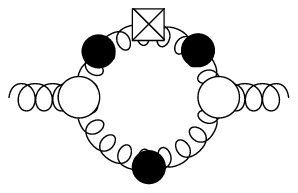}} 
{\large$\;-\;$}\parbox{50pt}{\includegraphics[width=\linewidth]{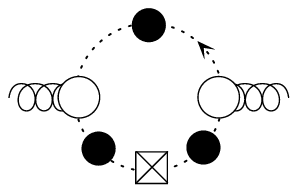}}  
{\large$\;-\;$}\parbox{50pt}{\includegraphics[width=\linewidth]{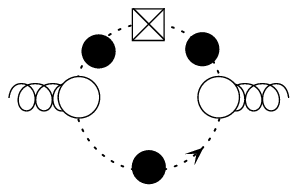}}  
{\large$\;-\;$}\parbox{35pt}{\includegraphics[width=\linewidth]{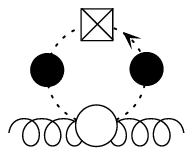}}  
{\large$\;-\fc{1}{2}\;$}\parbox{35pt}{\includegraphics[width=\linewidth]{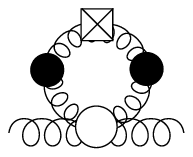}}  \\
{\large$k\partial_k$} \parbox{30pt}{\includegraphics[width=\linewidth]{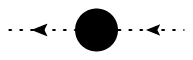}}{\large$^{-1}=$} 
\parbox{50pt}{\includegraphics[width=\linewidth]{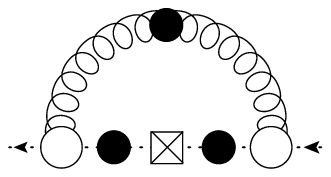}} 
{\large$\;+\;$}\parbox{50pt}{\includegraphics[width=\linewidth]{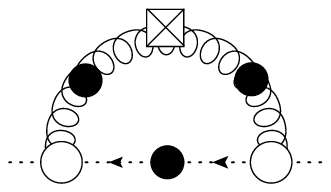}} 
{\large$\;-\fc{1}{2}\;$}\parbox{35pt}{\includegraphics[width=\linewidth]{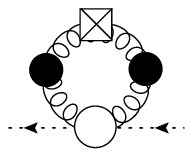}} 
{\large$\;-\;$}\parbox{35pt}{\includegraphics[width=\linewidth]{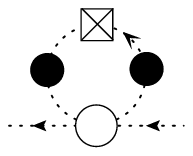}} 
\caption{Flow equations of the propagators. The spiral and dotted lines with black circles denote the regularized gluon and ghost propagators at cutoff momentum $k$, respectively. White circles stand for proper vertices at cutoff $k$, a regulator insertion $\dot{R}_k$ is represented by a square with a cross.} 
\label{full flow}
\end{figure*}

The FRG flow equations embody an infinite tower of coupled equations
for the flow of the propagators and the proper vertices. These
equations have to be truncated to get a closed system. We shall use
the following truncation: we only keep the gluon and ghost
propagators, to wit
\be
\label{25}
\Gamma^{(2)}_{k,AA}  =  2 \omega_k (p) \,, \qquad\qquad
\Gamma^{(2)}_{k,\bar{c} c}  =  \frac{p^2}{d_k (p)} \hk ,
\ee
In addition, we keep the ghost-gluon vertex $\Gamma^{(3)}_{k,A \bar{c}
  c}$, which we assume to be bare, i.e., we do not solve its
FRG flow equation. The above truncation removes the tadpole diagrams
from Fig. \ref{full flow}. Moreover, we shall assume infrared ghost
dominance and discard gluon loops. Then the flow equations of the
ghost and gluon propagator reduce to the ones shown in Fig. \ref{trunc
  flow}.

\begin{figure*}[h] 
\centering 
{\large$k\partial_k$} \parbox{30pt}{\includegraphics[width=\linewidth]{gluon-propagator.eps}}{\large$^{-1}=$} 
{\large$\;-\;$}\parbox{50pt}{\includegraphics[width=\linewidth]{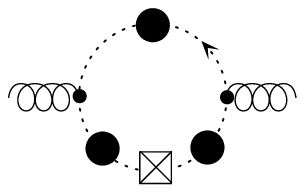}} 
{\large$\;-\;$}\parbox{50pt}{\includegraphics[width=\linewidth]{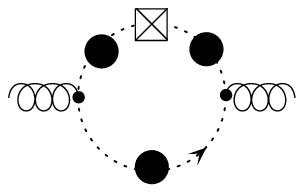}}{\large$\,;\;$} \qquad 
{\large$k\partial_k$} \parbox{30pt}{\includegraphics[width=\linewidth]{ghost-propagator.eps}}{\large$^{-1}=$}  
\parbox{50pt}{\includegraphics[width=\linewidth]{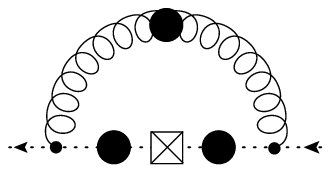}} 
{\large$\;+\;$}\parbox{50pt}{\includegraphics[width=\linewidth]{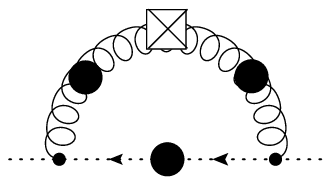}} 
\caption{Truncated propagator flow equations. The bare vertices at $k=\Lambda$ are symbolized by small dots.}  
\label{trunc flow}
\end{figure*}  

\begin{wrapfigure}{R}{0.47\linewidth}
\includegraphics[width=\linewidth]{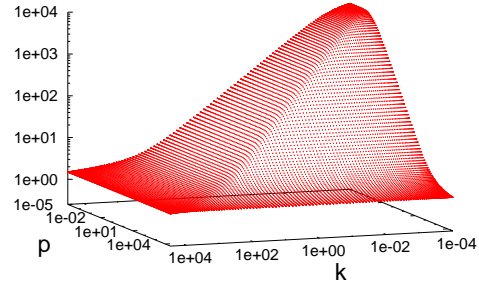}
\caption{Flow $d_k(p)$ of the ghost form factor.}
\label{ghost_flow_full.eps}
\end{wrapfigure}

\begin{figure*}[t]
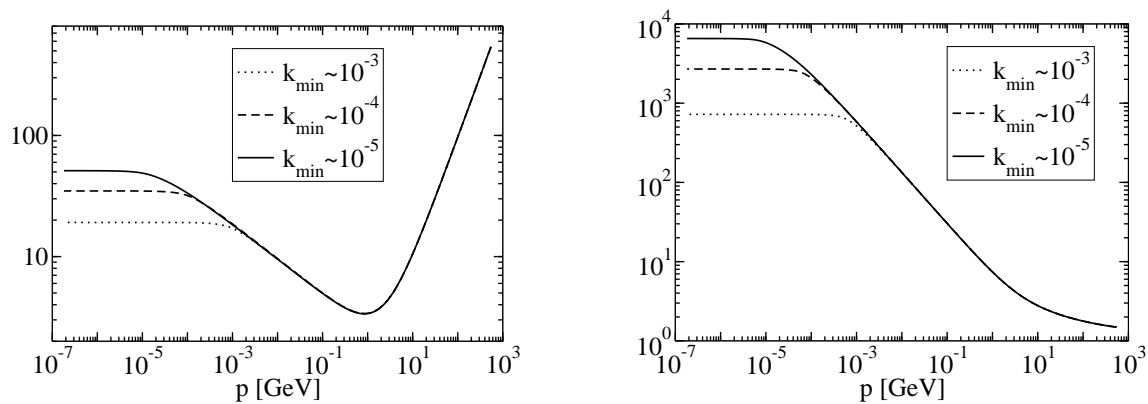
 
\centering
\includegraphics[width=0.45\linewidth]{3diff_kmin_omega.eps}
\hfill
\includegraphics[width=0.45\linewidth]{3diff_kmin_ghost.eps} 
\caption{Inverse gluon propagator $\omega$ and ghost form factor $d$ at three minimal cutoff values $k_{min}$.}  
\label{3diff_kmin_omega_ghost}
\end{figure*}

The flow equations in Fig. \ref{trunc flow}  are solved numerically using the regulators
\be\begin{split}
 & R_{A,k} (p) = 2p r_k (p) \;, \hk  R_{c,k} (p) = p^2 r_k (p) \hk , \\ & r_k (p) 
= \exp \left[ \frac{k^2}{p^2} - \frac{p^2}{k^2} \right]
\end{split}\ee
and the perturbative initial conditions at the large momentum scale $k = \Lambda$,
\be
\label{29}
d_{\Lambda} (p) = d_\Lambda = const.\;,\quad \omega_\Lambda (p) = p +
a \hk .  \ee With these initial conditions, the flow equations for the
ghost and gluon propagators are solved under the constraint of
infrared scaling for the ghost form factor. The resulting full flow of
the ghost dressing function is shown in
Fig. \ref{ghost_flow_full.eps}. As the IR cut-off momentum $k$ is
decreased, the ghost form factor $d_k(p)$ (constant at $k = \Lambda$)
builds up infrared strength and the final solution at $k = k_{min}$ is
shown in Fig. \ref{3diff_kmin_omega_ghost} together with the one for
the gluon energy $\omega_{k_{min}}(p)$.  It is seen that the IR
exponents, i.e., the slopes of the curves $d_{k_{min}} (p) ,
\omega_{k_{min}} (p)$ do not change as the minimal cut-off $k_{min}$
is lowered. Let us stress that we have assumed infrared scaling of the
ghost form factor but not the horizon condition. The latter was
obtained from the integration of the flow equation but not put in by
hand (the same is also true for the infrared analysis of the
Dyson-Schwinger equations following from the variational Hamiltonian
approach, i.e., assuming scaling the DSEs yield the horizon
condition).
The infrared exponents extracted from the numerical solutions of the flow equations are
\be
\label{31}
\alpha = 0.28 \, , \quad \beta = 0.64 \hk .  \ee They satisfy the sum
rule (\ref{8}) resulting from the DSE obtained in the variational
approach but are smaller than the ones of the DSE, see
Eq. (\ref{9}). Moreover, the present approach allows to prove the
uniqueness of the sum rule (\ref{8}) \cite{Leder:2010ji}, analogously
to the proof in Landau gauge \cite{Fischer:2009tn}.

Replacing the propagators with running cut-off momentum scale $k$
under the loop integrals of the flow equation by the propagators of
the full theory, \be
\label{32}
d_k (p) \to d_{k = 0} (p) \hk , \hk \omega_k (p) \to \omega_{k = 0}
(p) \hk , \ee amounts to taking into account the tadpole diagrams
\cite{Leder:2010ji}. Then the flow equations can be analytically
integrated and turn precisely into the DSEs obtained in the
variational approach to the Hamiltonian formulation of Yang-Mills
theory \cite{Feuchter:2004mk}, with explicit UV regularization by
subtraction. This establishes the connection between these two
approaches and highlights the inclusion of a consistent UV
renormalization procedure in the present approach.

The above results encourage further studies, which includes the
flow of the potential between static color sources as well as dynamic
quarks.

\end{document}